\documentclass[11pt]{article}

\usepackage{graphicx}
\usepackage{amsmath, amsthm, amssymb}
\usepackage{algorithm}
\usepackage{algorithmic}
\usepackage{subfigure}
\usepackage{comment}
\usepackage{url}

\newcommand{\R}{\mathbb{R}}

\newcommand{\Z}{\mathbb{Z}}

\newcommand{\F}{\mathbb{F}}

\newcommand{\ket}[1]{| #1 \rangle}

\newcommand{\be}{\begin{equation}}
\newcommand{\ee}{\end{equation}}
\newcommand{\bea}{\begin{eqnarray}}
\newcommand{\eea}{\end{eqnarray}}
\newcommand{\bes}{\begin{equation*}}
\newcommand{\ees}{\end{equation*}}
\newcommand{\beas}{\begin{eqnarray*}}
\newcommand{\eeas}{\end{eqnarray*}}

\newtheorem{thm}{Theorem}
\newtheorem*{thm*}{Theorem}
\newtheorem{cor}[thm]{Corollary}

\newtheorem{lem}[thm]{Lemma}
\newtheorem*{lem*}{Lemma}

\begin{document}

\date{\today}

\title{Quantum algorithms for shifted subset problems}

\author{Ashley Montanaro\footnote{montanar@cs.bris.ac.uk}\\ \\ {\small Department of Computer Science, University of Bristol,}\\{\small Woodland Road, Bristol, BS8 1UB, UK.} }

\maketitle

\begin{abstract}
We consider a recently proposed generalisation of the abelian hidden subgroup problem: the {\em shifted subset problem}. The problem is to determine a subset $S$ of some abelian group, given access to quantum states of the form $\ket{S+x}$, for some unknown shift $x$. We give quantum algorithms to find Hamming spheres and other subsets of the boolean cube $\{0,1\}^n$. The algorithms have time complexity polynomial in $n$ and give rise to exponential separations from classical computation.
\end{abstract}

\section{Introduction}

It is widely believed that certain problems can be solved significantly more quickly using a quantum computer than is possible classically. The canonical example of such a speed-up is Shor's factoring algorithm \cite{shor97}. In common with most known super-polynomial quantum speed-ups, the algorithm exploits a hidden structure in the input (in this case, hidden periodicity).

Shor's algorithm has been generalised to give a polynomial-time quantum algorithm for the {\em abelian hidden subgroup problem}, which can be defined as follows \cite{brassard97,jozsa97}. Given an abelian group $G$, and a function $f:G \rightarrow S$ (for some arbitrary set $S$) promised to be constant on cosets of some subgroup $H \le G$ and distinct on each coset, find a set of generators for $H$. One of the major open challenges in the field of quantum algorithms is the solution of the further generalisation of this problem to non-abelian groups. (Among other things, this would give an efficient quantum algorithm for the graph isomorphism problem \cite{beals97,jozsa01}.) However, there are some indications that the general non-abelian hidden subgroup problem might be hard for quantum computation as well \cite{hallgren06,moore05}.

With this in mind, it is natural to look for other hidden structures that can be exploited by a quantum computer, perhaps of a less group-theoretic nature. Notable examples of problems that are based on such structures are the generalised hidden shift problem of Childs and van Dam \cite{childs07b} and the hidden polynomial problem of Childs, Schulman and Vazirani \cite{childs07a}, a modification of which was also studied by Decker, Draisma and Wocjan \cite{decker07a,decker07}. This paper is concerned with another problem of this kind, which was also recently defined by Childs et al.\ in \cite{childs07a}: the {\em shifted subset problem}.

This problem can be informally stated as follows. Given a machine producing quantum states which are equal superpositions of the elements in some subset $S$ of an abelian group $G$, with each element shifted by an equal but unknown offset $x$, determine $S$. Given the promise that $S$ is picked from a certain family of subsets, the hope is to find an efficient quantum algorithm for this problem. If $S$ were promised to be a subgroup of $G$, this could be achieved using the quantum algorithm for the abelian hidden subgroup problem.

Of course, this formulation is intrinsically quantum, and we would like the problem to make sense classically too. The black-box approach taken for the abelian hidden subgroup problem will not be suitable, as shifted subsets may intersect. However, it turns out that one can define (see Section \ref{sec:blackbox}) an oracular problem corresponding to any shifted subset problem. This problem can be solved efficiently by a quantum algorithm, assuming that its related shifted subset problem can be, but requires exponential time to be solved classically.

The work of Childs et al.\ focused on the additive group of $\F_q^n$, for $\F_q$ a finite field with $q$ a prime power, and $n$ constant. In this work, we consider quantum algorithms to find hidden subsets of the boolean cube $\{0,1\}^n$. This is a natural generalisation of the problem used by Simon \cite{simon97} to show the first exponential separation between quantum and classical bounded-error computation, which is also a special case of the abelian hidden subgroup problem. (In the case of Simon's problem, the shifted subsets are lines.) Explicitly, the general formulation of the problem is as follows.

\begin{figure}[h]
\noindent \framebox{
\begin{minipage}{5.4truein}
{\sc Shifted Subset Problem}. An unknown subset $S \subseteq \{0,1\}^n$ is picked from a known family of subsets $\mathcal{F}$. Given an oracle that produces a supply of states of the form
\[ \ket{S + x} = \frac{1}{\sqrt{|S|}} \sum_{s \in S} \ket{s + x}, \]
for some unknown and varying shift $x$, determine $S$.
\end{minipage}
}
\end{figure}

We generally look for quantum algorithms that run in time polynomial in $n$. (Note that this is a different regime to that considered by Childs et al.\ in \cite{childs07a}, where the dimension $n$ was considered to be constant, and $q$ grew; in that paper, quantum algorithms were sought that ran in time $\mathrm{polylog}(q)$.) One particular class of subsets considered by Childs et al.\ were hidden spheres in $\F_q^n$ (a point $x=(x_1,...,x_n)$ is said to be on the sphere in $\F_q^n$ with radius $r \in \F_q$ centred at the origin if $\sum_i x_i^2=r$). Here, we consider the natural counterpart for the cube: spheres with respect to the Hamming weight. The formal definition of this problem is given below.

\begin{figure}[t]
\noindent \framebox{
\begin{minipage}{5.4truein}
{\sc Shifted Sphere Problem}. Let $S_r$ be the subset $\{x \in \{0,1\}^n | |x|=r\}$, where $|x|$ is the Hamming weight of $x$ and $0 \le r \le n/2$. Given an oracle that produces a supply of states of the form
\[ \ket{S_r + x} = \frac{1}{\sqrt{|S_r|}} \sum_{s \in S_r} \ket{s + x}, \]
for some unknown and varying shift $x$, determine $r$.
\end{minipage}
}
\end{figure}

The main contribution of this paper is an explicit polynomial-time quantum algorithm for the shifted sphere problem in any dimension $n$. This is in contrast to the algorithm of Childs et al., which only determined a single bit of $r$, and only for $n$ odd. (On the other hand, it was shown in \cite{childs07a} that the quantum query complexity of determining $r$ completely is polynomial when $n$ is odd, and evidence was presented for this being true for even $n$ as well.) Our algorithm's running time is unpalatably high ($O(n^4)$ for $n$ odd and $O(n^6)$ for $n$ even); however, in the case where $n$ is even, we give a linear-time quantum algorithm to determine whether $r$ is even or odd. We also give efficient and quite straightforward quantum algorithms for three other classes of shifted subset problems: sets of greatly different sizes, juntas and parity functions.

The quantum component of all the algorithms in this paper is extremely simple, consisting only of the {\em Fourier sampling} primitive of Bernstein and Vazirani \cite{bernstein97}. However, we do not see this as a disadvantage, but rather as an example of how classical postprocessing can unlock the power of a simple quantum circuit. This work also hints that there may exist other, more subtle problems for which this primitive may be of use. We note that Bshouty and Jackson~\cite{bshouty99}, and also Atici and Servedio \cite{atici07}, have proposed quantum algorithms that use only the Fourier sampling primitive for tasks in computational learning theory.

The paper is organised as follows. Section \ref{sec:algorithm} describes the generic quantum part of algorithms for shifted subset problems. Section \ref{sec:spheres} gives a quantum algorithm for the shifted sphere problem, and Section \ref{sec:other} describes other shifted subset problems for which we have efficient quantum algorithms. Section \ref{sec:blackbox} gives an oracular problem which can be used to show an exponential quantum-classical separation for shifted subset problems. Section \ref{sec:conclusion} has some brief concluding remarks.

We now proceed to give the general quantum algorithm for shifted subset problems.


\section{The quantum algorithm}
\label{sec:algorithm}

Assume that we wish to determine a secret subset $S \subseteq \{0,1\}^n$ using an oracle which provides us with a supply of states of the form
\[ \ket{S+x} = \frac{1}{\sqrt{|S|}} \sum_{y \in S} \ket{y + x} \]
where the shift $x$ is unknown. Call these {\em shifted subset states}. Our goal is to use these states to determine $S$. The quantum component of the algorithm for doing so is in fact independent of $S$. In order to remove the unknown and unwanted shift $x$, the first step is to perform the Fourier transform over $\Z_2^n$ by applying Hadamard gates to each qubit,
\[ H^{\otimes n} \ket{S+x} = \frac{1}{\sqrt{|S| 2^n}} \sum_{y \in S} \sum_{z \in \{0,1\}^n} (-1)^{z\cdot (y+x)} \ket{z} = \frac{1}{\sqrt{|S| 2^n}} \sum_{z \in \{0,1\}^n} (-1)^{x \cdot z} \sum_{y \in S} (-1)^{y\cdot z} \ket{z}. \]
Next, we measure this state, giving rise to the following probability distribution, which will be called $\pi_S$:
\be \label{eqn:fourierSample} \pi_S(z) = \frac{1}{|S| 2^n} \left( \sum_{y \in S} (-1)^{y\cdot z} \right)^2. \ee
We can now attempt to use samples from this distribution ({\em Fourier samples} \cite{bernstein97}), which are $n$-bit strings, to infer $S$. In Appendix B, we give a general upper bound on the number of Fourier samples required to identify shifted subsets (without considering the complexity of postprocessing the results). In order to understand this distribution, it will be useful to borrow some basic ideas from Fourier analysis; for notation, see \cite{deitmar05}. For a function $f:\{0,1\}^n \rightarrow \R$, we have the following notation for the Fourier transform of $f$ over $\Z_2^n$:
\[ \hat{f}(x) = \frac{1}{\sqrt{2^n}} \sum_{y \in \{0,1\}^n} (-1)^{x\cdot y} f(y). \]
This allows us to express $\pi_S$ in a compact way in terms of the Fourier transform of the characteristic function ${\bf 1}_S$,
\[ \pi_S = \frac{\widehat{{\bf 1}_S}^2}{|S|}. \]


\section{Shifted spheres}
\label{sec:spheres}

We will now consider a particular family of subsets: Hamming spheres. That is, the family of subsets $\{S_r\}$, where $x \in S_r$ if and only if $|x| = r$. $r$ is the {\em radius} of the sphere. Our goal will be to find the radius of such a sphere, given access to states of the form
\[ \ket{S_r+x} = \frac{1}{\sqrt{\binom{n}{r}}} \sum_{y \in \{0,1\}^n, |y|=r} \ket{y + x} \]
for some arbitrary shift $x$. (Note that we must insist that $r \le n/2$, as states of the form $\ket{S_r+x}$ are indistinguishable from those of the form $\ket{S_{n-r}+x'}$.) We give a bounded-error quantum algorithm to find $r$ in polynomial time.


\subsection{Preliminaries}

In a mild abuse of notation, define $\pi_r = \pi_{S_r}$. Then
\[ \pi_r(x) = \frac{1}{\binom{n}{r} 2^n} \left( \sum_{y \in \{0,1\}^n,|y|=r} (-1)^{x\cdot y}\right)^2, \]
where $x$ is an $n$-bit string. These sums have been much studied in the coding theory literature and are known as {\em Krawtchouk polynomials} \cite{krasikov99,macwilliams83}. The $r$'th Krawtchouk polynomial is defined as
\[ K_r^n(x) = \sum_{y\in \{0,1\}^n,|y|=r} (-1)^{x\cdot y} = \sum_{i=0}^r (-1)^i \binom{|x|}{i}\binom{n-|x|}{r-i}. \]
We use the notation $K_r(x)$ when $n$ is understood to be fixed. Thus we have $\pi_r(x) = K_r(|x|)^2/(\binom{n}{r} 2^n)$. As this distribution depends only on $|x|$, we will only consider the Hamming weight of the outcome and redefine $x = |x|$, $0 \le x \le n$, giving
\[ \pi_r(x) = \frac{\binom{n}{x} K_r(x)^2}{\binom{n}{r} 2^n}. \]
We will sample from this distribution to determine $r$.


\subsection{A polynomial-time quantum algorithm}

We now give a polynomial-time bounded-error quantum algorithm that determines the radius $r$ from samples of $\pi_r$. The algorithm runs in time $O(n^6)$ for $n$ even, and $O(n^4)$ for $n$ odd.

Our goal will be to estimate $r$ by estimating the probabilities of certain measurement outcomes which occur often. In particular, those with the Hamming weights $n/2$, $n/2-1$ and $n/2+1$ for $n$ even, and $(n-1)/2$ and $(n+1)/2$ for $n$ odd. We start by determining the probabilities of these outcomes. Note that the probability of any given outcome can be calculated efficiently, as there is an $O(n \log^2 n)$ algorithm to evaluate Krawtchouk polynomials~\cite{driscoll97}.

\begin{lem}
\label{lem:krawtprob}
If $n$ is even, then the probability of obtaining an outcome with weight $n/2$ is
\[ \pi_r(n/2) = \left\{ \begin{array}{ll}
\frac{1}{2^n} \binom{r}{r/2}\binom{n-r}{(n-r)/2} & \mbox{($r$ even)}\\
0 & \mbox{($r$ odd)} \end{array}\right., \]
and the probability of obtaining an outcome with weight $n/2-1$ or $n/2+1$ is
\[ \pi_r(n/2-1) + \pi_r(n/2+1) = \left\{ \begin{array}{ll}
\frac{(n-2r)^2}{2^{n-1} n(n+2)} \binom{r}{r/2}\binom{n-r}{(n-r)/2} & \mbox{($r$ even)}\\
\frac{r(n-r)}{2^{n-5} n(n+2)} \binom{r-1}{(r-1)/2}\binom{n-r-1}{(n-r-1)/2} & \mbox{($r$ odd).} \end{array}\right. \]
If $n$ is odd, then the probability of obtaining an outcome with weight $(n-1)/2$ or $(n+1)/2$ is
\[ \pi_r((n-1)/2) + \pi_r((n+1)/2) = \left\{ \begin{array}{ll}
\frac{n-r}{2^{n-2} (n+1)} \binom{r}{r/2}\binom{n-r-1}{(n-r-1)/2} & \mbox{($r$ even)}\\
\frac{r}{2^{n-2}(n+1)} \binom{r-1}{(r-1)/2}\binom{n-r}{(n-r)/2} & \mbox{($r$ odd).} \end{array}\right. \]
\end{lem}

\begin{proof}
Deferred to Appendix A.
\end{proof}

In the case of $n$ even, this immediately suggests the following algorithm to determine whether $r$ is even or odd. Sample a number of times from $\pi_r$, output that $r$ is even if any outcomes with weight $n/2$ are obtained, and output ``odd'' otherwise. Using the inequality that
\be \label{eqn:binomapprox} \frac{2^{2n}}{\sqrt{\pi n}} \ge \binom{2n}{n} \ge \frac{2^{2n}}{\sqrt{2 \pi n}}, \ee
which is valid for $n>0$, it is clear that for $r=0$, $\pi_r(n/2) = \Omega(1/\sqrt{n})$. For even $r>0$,
\be \label{eqn:stirling} \pi_r(n/2) \ge \frac{1}{\pi} \frac{1}{\sqrt{r(n-r)}} = \Omega(1/n), \ee
which implies that we need only $O(n)$ samples to determine the last bit of $r$ with constant probability of success.

We now describe a slightly more complex algorithm for calculating $r$ completely, again when $n$ is even. Take $k$ samples from $\pi_r$, for some $k$ to be determined. Count the number of times $t_1$ that an outcome with Hamming weight $n/2$ occurs, and the number of times $t_2$ that an outcome with Hamming weight $n/2-1$ or $n/2+1$ occurs.

If $t_1>0$, then $r$ is even. In this case, consider the measurement process as a biased coin that produces heads with probability $\pi_r(n/2)$ and tails with probability $1-\pi_r(n/2)$.  By a standard Chernoff bound argument, to estimate this probability within additive error $\epsilon$ with constant probability of success, it is sufficient to take the average of the number of heads obtained in $k=O(1/\epsilon^2)$ trials. To determine the number of trials required, we therefore need to lower bound the minimum, over all even $r \neq s$ with $r,s\le n/2$, of $|\pi_r(n/2)-\pi_s(n/2)|$. This can be carried out using Lemma \ref{lem:krawtprob}, as follows.

Consider the difference $\pi_r(n/2) - \pi_{r+2}(n/2)$, for $0 \le r \le n/2 - 2$. By expanding $\pi_{r+2}(n/2)$ in terms of $\pi_r(n/2)$, we have
\beas
\pi_r(n/2) - \pi_{r+2}(n/2) &=& \frac{1}{2^n} \left(1 - \frac{(n-r)(r+1)}{(n-r-1)(r+2)}\right) \binom{r}{r/2}\binom{n-r}{(n-r)/2}\\
&=& \frac{1}{2^n} \left( \frac{n-2(r+1)}{(n-r-1)(r+2)} \right) \binom{r}{r/2}\binom{n-r}{(n-r)/2}.
\eeas
The product of binomial coefficients can be lower bounded using eqn.\ (\ref{eqn:binomapprox}), and as $r \le n/2-2$, it is easy to see that the remaining fraction is positive and $\Omega(1/n^2)$. This implies that $|\pi_r(n/2) - \pi_s(n/2)| = \Omega(1/n^3)$ for all even $r\neq s$ with $r,s\le n/2$.
This in turn implies that we can estimate $\pi_r(n/2)$ -- and hence $r$ -- in the case where $r$ is even after a somewhat unappetising $k=O(n^6)$ trials.

On the other hand, consider the case where $t_1=0$ after $\Theta(n)$ trials. Hence we can assume that $r$ is odd, and estimate $p(r) \equiv \pi_r(n/2-1)+\pi_r(n/2+1)$ from $t_2$, using the same technique as in the previous paragraph. We calculate $p(r+2) - p(r)$ by expanding binomial coefficients:
\beas
p(r+2) - p(r)\!\!\!\!&=&\!\!\!\!\frac{1}{2^{n-5}} \frac{r}{n(n+2)} \left(\frac{(r+2)(n-r-1)}{r+1}\!-\!(n-r)\right)\!\binom{r-1}{(r-1)/2}\!\binom{n-r-1}{(n-r-1)/2}\\
&=&\!\!\!\!\frac{1}{2^{n-5}} \frac{r(n-2(r+1))}{n(n+2)(r+1)} \binom{r-1}{(r-1)/2}\binom{n-r-1}{(n-r-1)/2}.
\eeas
Using the approximation (\ref{eqn:binomapprox}) gives the result that, for all $r,s<n/2$ with $r \neq s$, $|p(r) - p(s)| = \Omega(n^{-3})$, implying that we can calculate the hidden radius in this case with a constant probability of success after $O(n^6)$ trials as well.

We finally turn to the case of $n$ odd. Here, we consider the measurement outcomes with weight $(n-1)/2$ and $(n+1)/2$. Again, we treat the measurement process as a biased coin, and attempt to estimate $p'(r) \equiv \pi_r((n-1)/2)+\pi_r((n+1)/2)$. First, note that the probability $p'(r)$ for $r$ even is equal to $p'(n-r)$, where $n-r$ is now odd. As before, we can calculate $p'(r+2) - p'(r)$ for $r$ odd by expanding binomial coefficients:
\beas
p'(r+2) - p'(r) &=& \frac{1}{2^{n-2}} \frac{r}{n+1} \left(\frac{(r+2)(n-r)}{(r+1)(n-r-1)} - 1\right) \binom{r-1}{(r-1)/2}\binom{n-r}{(n-r)/2}\\
&=& \frac{1}{2^{n-2}} \frac{r}{(r+1)(n-r-1)} \binom{r-1}{(r-1)/2}\binom{n-r}{(n-r)/2}.
\eeas
The product of binomial coefficients divided by $2^{n-2}$ can be estimated as $\Omega(1/n)$, implying $p'(r+2) - p'(r) = \Omega(n^{-2})$. Using the Chernoff bound argument again, in order to infer $r$ with a constant probability of error by estimating the probability $\pi_r((n-1)/2) + \pi_r((n+1)/2)$, $O(n^4)$ trials are sufficient.


\subsection{Shifted Hamming balls}

We make a brief remark regarding the related structure of balls with respect to the Hamming weight, i.e.\ the family of subsets $\{B_r\}$, where $x \in B_r$ if and only if $|x| \le r$. $r$ is the {\em radius} of the ball. The shifted subset problem for Hamming balls is to find the radius of such a ball, given access to states of the form
\[ \ket{B_r+x} = \frac{1}{\sqrt{\sum_{k=0}^r \binom{n}{k}}} \sum_{y \in \{0,1\}^n, |y|\le r} \ket{y + x} \]
for some arbitrary shift $x$. In this case, it is not necessary to insist that $r \le n/2$ for the problem to be well-defined.

As a ball is simply the union of disjoint spheres, its Fourier transform is a sum of Krawtchouk polynomials. This can be simplified using a convenient expression for the sum of Krawtchouk polynomials \cite{macwilliams83},
\[ \sum_{s=0}^r K_s^n(x) = K_r^{n-1}(x-1). \]
This implies that the probability distribution $\pi_r(x)$ that gives the probability of obtaining an outcome with weight $x$ when the hidden ball has radius $r$ is given by
\[ \frac{\binom{n}{x} K_r^{n-1}(x-1)^2}{2^n \sum_{k=0}^r \binom{n}{k}}. \]
Similar results can thus be obtained for finding shifted balls as in the case of shifted spheres.


\section{Other shifted subsets}
\label{sec:other}

In this section, we note three further classes of shifted subsets which can be distinguished efficiently by a quantum algorithm.

\subsection{Subsets that differ greatly in size}
\label{subsection:size}

For a particular subset $S$, consider the probability of obtaining the zero string from a Fourier sample, $\pi_S(0^n)$. It can be verified directly from eqn.\ (\ref{eqn:fourierSample}) that this is equal to $|S|/2^n$. Consider an unknown subset $S$ picked from a family of subsets $\{S_i\}$ which all have significantly different sizes, i.e. $\left| |S_i| - |S_j| \right| = 2^n/\mathrm{poly}(n)$ for $i \neq j$. Then $|S|$ can be estimated, and thus $S$ determined, from a polynomial number of Fourier samples simply by counting the number of zero weight outcomes received, and using this to estimate $\pi_S(0^n)$.

\subsection{Juntas}

Consider a set of non-constant boolean functions $\{f_1,\dots,f_k\}$ on $n$ variables that each depend on a different, disjoint subset $T_k \subset [n]$ of variables, where each subset is of constant size. Such functions are known as {\em juntas}. Define a family of subsets $\mathcal{F} = \{S_k\}$ by letting $f_k$ be the characteristic function of $S_k$. As $f_k$ is not constant, $|S_k| \le 2^{n-|T_k|}(2^{|T_k|}-1) = c 2^n$ for some constant $c<1$.

It is easy to see that the Fourier transform of the shifted subset state corresponding to $S_k$ has no support on those bit-strings which have 1's in positions outside of $T_k$. (This is just saying that the Fourier transform of a function that depends on a certain set of variables depends only on those same variables.) This implies that, as the sets $\{T_k\}$ are all disjoint, obtaining any measurement outcome other than the zero string is sufficient to identify $T_k$, and thus $S_k$. However, as noted in Section \ref{subsection:size}, $\pi_{S_k}(0^n) = |S_k|/2^n = c<1$, so a constant number of repetitions suffice to obtain such a non-zero measurement outcome.

\subsection{Parity functions}

Consider a subset $S \subseteq \{0,1\}^n$ consisting of those bit-strings whose parity on some non-empty subset $T$ of the $n$ bits is 1. That is, the characteristic function ${\bf 1}_S(x) = \bigoplus_{i \in T} x_i$. Note that the set $S$ is not a subgroup of $\{0,1\}^n$, although its complement is a subgroup.

Let $t$ be the bit-string corresponding to the characteristic function ${\bf 1}_T$ (that is, $t_i=1 \Leftrightarrow i \in T$). Then it is easy to see that the Fourier transform $\widehat{{\bf 1}_S}$ is supported on only two positions:
\beas
\widehat{{\bf 1}_S}(x) &=& \frac{1}{\sqrt{2^n}} \left( \sum_{y \in \{0,1\}^n} (-1)^{x \cdot y} (1-(-1)^{t \cdot y})/2 \right)\\
&=& \frac{1}{\sqrt{2^{n+2}}} \left( \sum_{y \in \{0,1\}^n} (-1)^{x \cdot y} - \sum_{y \in \{0,1\}^n} (-1)^{(x+t) \cdot y} \right),
\eeas
which is zero unless $x$ is equal to $t$ or the zero string; moreover, in both these cases $|\widehat{{\bf 1}_S}(x)|$ is equal. Therefore, sampling from the distribution $\pi_S$ will give $t$ with probability $1/2$, so only a constant number of samples is necessary to determine $t$, and thus $S$, with any desired constant probability.

We can generalise this family of subsets as follows, again in terms of characteristic functions. Let $T$ be a subset of the first $k$ bits, for some arbitrary $k$. Consider a family of $2^k$ subsets $\mathcal{F}$ for which, for each $S \in \mathcal{F}$, the characteristic function ${\bf 1}_S(x) = \bigoplus_{i \in T} x_i \oplus f_S(x_{k+1},\dots,x_n)$ for a distinct $T$, where $f_S$ is an arbitrary boolean function of the remaining $n-k$ bits. Thus ${\bf 1}_S$ is the sum of the bits in $T$, added to some arbitrary function of the remaining bits. Again, let $t$ be the $k$-bit string corresponding to ${\bf 1}_T$. Now the Fourier transform of ${\bf 1}_S$ will have support only on the zero bit-string and those bit-strings whose first $k$ bits are equal to $t$:
\beas
\widehat{{\bf 1}_S}(x) &=& \frac{1}{\sqrt{2^n}} \left( \sum_{y \in \{0,1\}^n} (-1)^{x \cdot y} (1-(-1)^{t \cdot y_{[k]} + f_S(y_{k+1},\dots,y_n)})/2 \right) \\
&=& \frac{1}{\sqrt{2^{n+2}}} \left( \sum_{y \in \{0,1\}^n} (-1)^{x \cdot y} - \sum_{y \in \{0,1\}^{n-k}} (-1)^{f_S(y)+x_{-[n-k]} \cdot y} \sum_{z \in \{0,1\}^k} (-1)^{(x_{[k]}+t) \cdot z} \right),
\eeas
where the notation $x_{[k]}$ represents the first $k$ bits of $x$ and $x_{-[n-k]}$ represents the last $n-k$ bits of $x$. It is easy to see that this is zero unless $x=0$ or $x_{[k]}=t$. So, once more, a constant number of samples suffice to determine $S$.


\section{Black-box formulation of shifted subset problems}
\label{sec:blackbox}

In order to show separations between quantum and classical computation for shifted subset problems, we need to introduce an oracular problem based on shifted subset problems which is hard for classical computers to solve. The standard approach used for hidden subgroup problems is to define an oracle function which is constant on subset elements with a given shift, and distinct on each shift \cite{simon97}. This is not suitable for the generalised problem considered in this paper, as shifted subsets may intersect.

We therefore take a more general approach, which is very similar to that in Appendix A of \cite{childs07a}, but is modified to work for subsets of arbitrary size (the approach in \cite{childs07a} required subset sizes to be close). The black-box problem we define would make sense for arbitrary underlying abelian groups, but for concreteness we consider only $\{0,1\}^n$.

Consider a subset $S \subseteq \{0,1\}^n$. Order the elements of $S$ arbitrarily and let $S[x]$ be the $x$'th element of $S$. We will define the following three functions on the set $\{0,1\}^{2n}$ (this increase of input size from $n$ to $2n$ bits is needed to show classical hardness).

\begin{itemize}
\item A ``colouring'' operator $c: \{0,1\}^{2n} \rightarrow [2^{2n}]$, which maps each point on the $2n$-dimensional cube to one of $\lceil 2^{2n}/|S| \rceil$ ``colours'', i.e.\ integers between 1 and $2^{2n}$. We restrict $c$ to send exactly $|S|$ points to each colour, assuming that $|S|$ divides $2^{2n}$ for simplicity (if it does not, then send $|S|$ points to all colours but one, and give the rest of the points the remaining colour).

\item A ``shifting'' operator $s: \{0,1\}^{2n} \times [2^{2n}] \rightarrow \{0,1\}^n$. For each $c_0$, and for all $x$ such that $c(x)=c_0$, $s(x,c_0)=S[\pi_{c_0}(x)]+\sigma(c_0)$, where $\pi_{c_0}$ is an arbitrary bijection between the set of points with colour $c_0$ and $S$, and $\sigma$ is an arbitrary mapping between the set of colours used by $c$ and $\{0,1\}^n$. That is, $s$ maps $x$ to a point in $S$, translated by an unknown shift $\sigma(c_0)$. For $x$ such that $f(x)\neq c_0$, $s(x,c_0)$ returns an arbitrary point in $\{0,1\}^n$.

\item An ``uncolouring'' operator $c^{-1}:[2^{2n}] \times \{0,1\}^n \rightarrow \{0,1\}^{2n}$ which performs the operation $(c_0,s(x,c_0)) \mapsto x$ for a valid pair $(c_0,s(x,c_0))$, and maps all other arguments to arbitrary $2n$-bit strings.
\end{itemize}

An instance of the oracular problem is given by the 4-tuple $(S, c, s, c^{-1})$. Our task is to determine $S$, given access to these oracles. Given a complete set of points of a given colour, a classical algorithm can find $S$ by simply querying $s$ on those points. In fact, it may not even be necessary to query $s$ at all: determining the number of colours used by $c$ is sufficient to determine $|S|$, for example.

However, we now sketch a proof that, in the worst case, any classical algorithm must make many queries to $c$ to get any information about $S$. Let $c$ be picked uniformly at random from the set of valid colouring operators, and consider a deterministic algorithm that makes a sequence of queries to $c$. As $|S|\le 2^n$, at least $2^n$ colours are used. Because $c$ is random, by a standard birthday argument the algorithm must make $\Omega(2^{n/2})$ queries before it finds two points $x$, $y$ that have the same colour. But querying $s$ on two points that have different colours does not give any information about $S$.

On the other hand, consider the following quantum procedure. Query $c$ on a uniform superposition of all points $x \in \{0,1\}^{2n}$. Measure the output register, thus giving a superposition of all points with a particular colour $c_0$. Query $s$, then uncompute $c$. This will leave a superposition of items in $S$, offset by an unknown shift (that depends on $c_0$). In other words, we perform the following operations (ignoring normalisation factors).
\beas
\sum_{x \in \{0,1\}^{2n}} \ket{x} &\mapsto& \sum_{x \in \{0,1\}^{2n}} \ket{x}\ket{c(x)}\\
&\mapsto& \sum_{x,c(x)=c_0} \ket{x}\ket{c_0}\ket{S[\pi_{c_0}(x)] + \sigma(c_0)}\\
&\mapsto& \sum_{x,c(x)=c_0} \ket{0}\ket{c_0}\ket{S[\pi_{c_0}(x)] + \sigma(c_0)}.
\eeas
Ignoring the first two registers, this is a shifted subset state $\ket{S + \sigma(c_0)}$. Thus a polynomial-time quantum algorithm that obtains any information about the shifted subset $S$ implies an exponential separation between quantum and classical computation.


\section{Conclusion}
\label{sec:conclusion}

We have given polynomial-time quantum algorithms for finding shifted subset structures on the boolean cube, and in particular for determining the radius of a shifted sphere. By defining a suitable oracle problem, these lead to an exponential separation between quantum and classical computation.

There are a couple of obvious possible extensions of this work. Firstly, finding other interesting families of subsets of the cube for which we can solve the shifted subset problem, or indeed more efficient algorithms for the shifted sphere problem. Ideas from the field of analysis of boolean functions might be helpful here. Secondly, and more challengingly, finding a practical and non-oracular problem for which the techniques of this paper are of use remains an open problem.

The following two additional open questions were suggested by Andrew Childs. Firstly, could the results of this paper be generalised to the underlying group $\Z_k^n$, where $k$ is constant? And secondly, another problem considered in \cite{childs07a} was the {\em hidden flat of centres problem} in $\F_q^n$ (for $q$ large and $n$ constant). In this case, spheres are of constant radius, but their centres are constrained to lie in an affine subspace. The problem is to determine this subspace. Could an analogue of this problem be solved on the boolean cube?


\section*{Acknowledgements}

This work was supported by the EC-FP6-STREP network QICS. I would like to thank Richard Low and Michael Bremner for helpful discussions on the subject of this paper. I would also like to thank Andrew Childs for helpful comments on a previous version, and an anonymous referee for many helpful suggestions.

\appendix

\section*{Appendices}

\section{Proof of Lemma \ref{lem:krawtprob}}
\label{appendix:krawtchouk}

\noindent For the reader's convenience, we repeat the statement of Lemma \ref{lem:krawtprob}.

\begin{lem*}
If $n$ is even, then the probability of obtaining an outcome with weight $n/2$ is
\[ \pi_r(n/2) = \left\{ \begin{array}{ll}
\frac{1}{2^n} \binom{r}{r/2}\binom{n-r}{(n-r)/2} & \mbox{($r$ even)}\\
0 & \mbox{($r$ odd)} \end{array}\right., \]
and the probability of obtaining an outcome with weight $n/2-1$ or $n/2+1$ is
\[ \pi_r(n/2-1) + \pi_r(n/2+1) = \left\{ \begin{array}{ll}
\frac{(n-2r)^2}{2^{n-1} n(n+2)} \binom{r}{r/2}\binom{n-r}{(n-r)/2} & \mbox{($r$ even)}\\
\frac{r(n-r)}{2^{n-5} n(n+2)} \binom{r-1}{(r-1)/2}\binom{n-r-1}{(n-r-1)/2} & \mbox{($r$ odd).} \end{array}\right. \]
If $n$ is odd, then the probability of obtaining an outcome with weight $(n-1)/2$ or $(n+1)/2$ is
\[ \pi_r((n-1)/2) + \pi_r((n+1)/2) = \left\{ \begin{array}{ll}
\frac{n-r}{2^{n-2} (n+1)} \binom{r}{r/2}\binom{n-r-1}{(n-r-1)/2} & \mbox{($r$ even)}\\
\frac{r}{2^{n-2}(n+1)} \binom{r-1}{(r-1)/2}\binom{n-r}{(n-r)/2} & \mbox{($r$ odd).} \end{array}\right. \]
\end{lem*}

\begin{proof}
The proof uses several recurrences and other properties satisfied by the Krawtchouk polynomials \cite{krasikov99}. Firstly, for $n$ even, the case $\pi_r(n/2)$ can be calculated explicitly using the generating function representation of the Krawtchouk polynomials \cite{krasikov99}.
\[ \sum_{r=0}^\infty K_r(x) z^r = (1-z)^x (1+z)^{n-x}, \]
implying
\[ \sum_{r=0}^\infty K_r(n/2) z^r = (1-z)^{n/2} (1+z)^{n/2} = (1-z^2)^{n/2} = \sum_k (-1)^k \binom{n/2}{k} z^{2k}. \]
Therefore, for odd $r$, $K_r(n/2)$ and hence $\pi_r(n/2)$ are zero. For even $r$, we have $K_r(n/2) = (-1)^{r/2} \binom{n/2}{r/2}$, implying
\[ \pi_r(n/2) = \frac{\binom{n}{n/2}\binom{n/2}{r/2}^2}{\binom{n}{r}2^n} = \frac{1}{2^n} \binom{r}{r/2}\binom{n-r}{(n-r)/2}, \]
where the second equality follows by expanding the binomial coefficients and simplifying.

We will use the following recurrence to calculate $K_r(n/2-1)$ for $r$ even \cite{krasikov99}:
\[ x K_r(x-1) = (n-2r)K_r(x)-(n-x)K_r(x+1). \]
As $K_r(x)=(-1)^r K_r(n-x)$, this implies that
\[ K_r(n/2-1) = (1-2r/n) K_r(n/2) = (1-2r/n) (-1)^{r/2} \binom{n/2}{r/2}\;\mbox{($r$ even)}. \]
To find $K_r(n/2-1)$ for $r$ odd, we use another recurrence,
\[ K_{r}(x-1) = K_{r-1}(x)+K_{r-1}(x-1)+K_r(x), \]
from which, by substituting in the previously obtained value of $K_{r-1}(n/2-1)$, using the fact that $K_r(n/2)=0$ and simplifying, we obtain
\[ K_r(n/2-1) = 2(-1)^{(r-1)/2} \binom{n/2-1}{(r-1)/2}\;\mbox{($r$ odd)}. \]
These expressions give $\pi_r(n/2-1)=\pi_r(n/2+1)$ for $n$ even, following another expansion and simplification of binomial coefficients. In the case of odd $n$, we use yet another recurrence satisfied by the Krawtchouk polynomials,
\[ K_r^n(x) = K_r^{n-1}(x) + K_{r-1}^{n-1}(x). \]
One can easily verify that this gives the claimed expressions for $\pi_r((n-1)/2) = \pi_r((n+1)/2)$.
\end{proof}


\section{Upper bounds on query complexity}
\label{appendix:queries}

\noindent We would like to show that, for some particular family $\mathcal{F}=\{S_i\}$ of subsets, a subset $S$ picked from $\mathcal{F}$ can be determined using only a small number of ``Fourier samples'' from the probability distribution $\pi_S$. It turns out that it is sufficient to show that the distributions $\{\pi_{S_i}\}$ are far apart in a pairwise sense \cite{barnum02,harrow06}.

We make the following definitions. The $p$-norm of a vector $f$ is $\|f\|_p = \left(\sum_x |f_x|^p\right)^{1/p}$, and the (Schatten) $p$-norm of a matrix $\rho$ is $\|\rho\|_p = \|\sigma(\rho)\|_p$, where $\sigma(\rho)$ is the vector of singular values of $\rho$. The {\em fidelity} of two quantum states $\rho_i$, $\rho_j$ is $F(\rho_i,\rho_j) = \|\sqrt{\rho_i}\sqrt{\rho_j}\|_1^2$. Then the following holds \cite{barnum02,harrow06}.

\begin{thm}
\label{thm:copies}
Let $S=\{\rho_i\}$ be a set of $N$ quantum states with $F(\rho_i,\rho_j)\le F$ for all $i \neq j$. Then, given $n$ copies of an unknown state $\rho_?$ from $S$, there is a measurement which determines the identity of $\rho_?$ with probability $1-\epsilon$ if
\[ n \ge \frac{2(\log N/\epsilon)}{\log 1/F} \]
\end{thm}
The following corollary is essentially immediate from a well-known inequality relating fidelity and trace distance, i.e.\ that $F(\rho_i,\rho_j) \le 1 - \frac{1}{4}\|\rho_i-\rho_j\|_1^2$.

\begin{cor}
\label{cor:copies}
Let $S=\{\rho_i\}$ be a set of $N$ quantum states with $\|\rho_i-\rho_j\|_1 \ge T$ for all $i \neq j$. Then, in order to determine which state an unknown state $\rho_?$ from $S$ is with constant probability of error, it is sufficient to have $n = O\left((\log N)/T^2\right)$ copies of $\rho_?$.
\end{cor}

Note that although this result holds for general quantum states, here it is applied only to classical probability distributions.

With Corollary \ref{cor:copies} in mind, we give an inequality that lower bounds the trace distance (aka $\ell_1$ distance) between two such probability distributions in terms of the characteristic functions of their sets. The inequality is based on the simplest case of the Hausdorff-Young inequality, which may be written down as $\|f\|_\infty \le \|\hat{f}\|_1/\sqrt{2^n}$. To state the inequality, we define the convolution operator $\ast$ as
\[ \label{eqn:convolution} (f \ast g)(x) = \frac{1}{\sqrt{2^n}} \sum_{y \in \{0,1\}^n} f(y) g(x + y). \]
It is a basic fact in Fourier analysis that the Fourier transform changes convolution into multiplication: $\widehat{f \ast g} = \hat{f} \hat{g}$ \cite{deitmar05}.

\begin{lem}
\label{lem:traceDistance}
Let $\pi_S$, $\pi_T$ be the probability distributions corresponding to the sets $S$, $T$. Then
\[ \| \pi_S - \pi_T \|_1 \ge \sqrt{2^n} \left\| \frac{{\bf 1}_S \ast {\bf 1}_S}{|S|} - \frac{{\bf 1}_T \ast {\bf 1}_T}{|T|} \right\|_\infty. \]
\end{lem}

\begin{proof}
By the Hausdorff-Young inequality, we have
\[ \sqrt{2^n} \left\| \frac{{\bf 1}_S \ast {\bf 1}_S}{|S|} - \frac{{\bf 1}_T \ast {\bf 1}_T}{|T|} \right\|_\infty \le \left\| \widehat{ \frac{{\bf 1}_S \ast {\bf 1}_S}{|S|}} - \widehat{\frac{{\bf 1}_T \ast {\bf 1}_T}{|T|} } \right\|_1 = \left\| \frac{{\widehat{{\bf 1}_S}}^2}{|S|} - \frac{{\widehat{{\bf 1}_T}}^2}{|T|} \right\|_1  = \| \pi_S - \pi_T \|_1. \]
\end{proof}
This inequality implies that the number of Fourier samples required to solve a shifted subset problem can be upper-bounded without needing to work directly with the Fourier transform of a set's characteristic function (which may be complicated). Unfortunately, when applied to the shifted sphere problem, the upper bound produced is no better than the time complexity of the algorithm of Section \ref{sec:spheres}.



\end{document}